\documentclass[showpacs,aps,prd,floatfix,amsmath,amssymb,nofootinbib]{revtex4-2}

\usepackage{graphicx}
\usepackage[utf8]{inputenc}
\usepackage[T1]{fontenc}
\usepackage{color}
\usepackage{epstopdf}
\usepackage{amssymb}
\usepackage{amsmath}
\usepackage{slashed}
\usepackage{float}
\makeatother
\begin{document}

\title{Chromomagnetic and chromoelectric dipole moments of quarks in the reduced 331 model}
\author{A. I. Hern\'andez-Ju\'arez$^\ast$}
\affiliation{Facultad de Ciencias F\'isico Matem\'aticas, Benem\'erita Universidad Aut\'onoma de Puebla, Apartado Postal 1152, Puebla, Puebla, M\'exico}
\email{alaban7_3@hotmail.com}
\author{A. Moyotl}
\affiliation{Ingenier\'ia en Mecatr\'onica, Universidad Polit\'ecnica de Puebla, Tercer Carril del Ejido Serrano s/n, San Mateo Cuanal\'a, Juan C. Bonilla, Puebla, Puebla, M\'exico}
\author{G. Tavares-Velasco}
\affiliation{Facultad de Ciencias F\'isico Matem\'aticas, Benem\'erita Universidad Aut\'onoma de Puebla, Apartado Postal 1152, Puebla, Puebla, M\'exico}
\date{\today}

\begin{abstract}
The one-loop contributions to the chromomagnetic dipole moment $\hat\mu_t(q^2)$ and electric dipole moment $\hat d_t(q^2)$ of the top quark are calculated within the reduced 331 model (RM331) for non-zero $q^2$. It is argued that the results are gauge independent  and thus represent valid observable quantities. In the RM331 $\hat \mu_t(q^2)$  receives new contributions from two heavy gauge bosons $Z'$ and $V^\pm$ and one neutral scalar boson $h_2$, along with a new contribution from the standard model Higgs boson via flavor changing neutral currents. The latter, which are also mediated by the $Z'$ gauge boson and the scalar boson $h_2$, can give a non-vanishing $\hat d_t(q^2)$ provided that there is a $CP$-violating phase.
The analytical results are presented in terms of both Feynman parameter integrals and Passarino-Veltman scalar functions, which are useful to cross-check the numerical results. Both $\hat\mu_t(q^2)$ and $\hat d_t(q^2)$ are numerically evaluated  for parameter values still allowed by the constraints from experimental data. It is found that the new one-loop contributions of the RM331 to the real (imaginary) part of $\hat \mu_t(q^2)$ are of the order of  $10^{-5}$ ($10^{-6}$), which are at least three orders of magnitude smaller than the standard model prediction, but are larger than the predictions of other models of new physics. In the RM331 the dominant contribution arising from the $V^\pm$ gauge boson for $\|q\|$ in the  30-1000 GeV interval and a mass  $m_{V}$ of the order of a few hundreds of GeVs. As for $\hat d_t(q^2)$, it receives its largest contribution from   $h_2$ exchange and can reach values of the order of $10^{-19}$, which is smaller than the contributions predicted by other standard model extensions.

\end{abstract}


\date{\today}

\maketitle

\section{Introduction}
The anomalous magnetic dipole moment (MDM) and the electric dipole moment are among the lepton properties that have stirred more interest in the experimental and theoretical areas. Currently, there is a discrepancy between the theoretical standard model (SM) prediction of the muon anomalous MDM and  its experimental measurement, which might be a hint of new physics \cite{ParticleDataGroup:2020ssz}. On the other hand, any experimental evidence of an electric dipole moment would give a clear signal of new sources of $CP$ violation as the SM contributions are negligibly small. With the advent of the LHC,  anomalous contributions  to the $\bar{t}tg$ coupling have also become the focus of interest. In analogy with the lepton electromagnetic vertex $\bar\ell\ell\gamma$,  the anomalous $\bar{q}qg$ coupling can be written as
\begin{equation}
\label{lag}
\mathcal{L}=-\frac{1}{2} \bar{q}\sigma^{\mu\nu} \left(\tilde{\mu}_q+i \tilde{d}_q \gamma ^5  \right)T^{a}q G_{\mu \nu}^a,
\end{equation}
where $\tilde{\mu}_q$ is the quark chromomagnetic dipole moment (CMDM) and  $\tilde{d}_q$ is the quark chromoelectric dipole moment (CEDM), whereas   $G^{\mu\upsilon}_a$ is the gluon field tensor and $T^a$ are the $SU(3)$ color generators.
It is also customary to define the CMDM and CEDM  in the dimensionless form  \cite{CMS:2016piu}
\begin{align}
\hat{\mu}_q \equiv\frac{m_q}{g_S} \widetilde{\mu}_q, \\
\hat{d}_q\equiv \frac{m_q}{g_S}\widetilde{d}_q.
\end{align}

On the experimental side, the search for  evidences of the anomalous top quark coupling $\bar{t}tg$  is underway at the LHC \cite{CMS:2019zct,CMS:2016piu,CMS:2019nrx}: the most recent   bounds  on the top quark CMDM and CEDM were obtained by the CMS collaboration \cite{CMS:2019nrx,CMS:2019kzp}, which managed to improve the previous bounds \cite{CMS:2016piu}  by one order of magnitude. Thus, one would expect that  more tight constraints on $\hat \mu_t$ and $\hat d_t$ could be set in the near future.

 As far as the theoretical predictions are concerned, in the SM the CMDM is induced at the one-loop level or higher orders via electroweak (EW) and  QCD contributions, whereas the CEDM can only arise  up to the three-loop level \cite{Shabalin:1978rs,Czarnecki:1997bu,Khriplovich:1985jr}.
The SM contributions to the on-shell $\hat\mu_t$ have already been studied in \cite{Martinez:2007qf,Choudhury:2014lna,Aranda:2018zis}, and more recently the scenario with an off-shell gluon was studied  in \cite{Hernandez-Juarez:2020drn,Aranda:2020tox} to address some ambiguities of previous calculations, particularly about the on-shell CMDM, which is divergent and meaningless in perturbative QCD. Since  both the top quark CMDM and CEDM could receive a considerable  enhancement from new physics contributions,  several calculations  have been reported in the literature within the framework of extension theories such as two-Higgs doublet models (THDMs) \cite{Gaitan:2015aia}, the four-generation THDM  \cite{Hernandez-Juarez:2018uow}, models with a heavy $Z'$ gauge boson \cite{Aranda:2018zis}, little Higgs models \cite{Cao:2008qd, Ding:2008nh}, the minimal supersymmetric standard model (MSSM) \cite{Aboubrahim:2015zpa},  unparticle models \cite{Martinez:2008hm}, vector like multiplet models \cite{Ibrahim:2011im}, etc. In this work we are interested in the contributions to the top quark CMDM and CEDM in  the reduced 331 model \cite{Ferreira:2011hm}.

The study of elementary particle models based on the $SU(3)_L \times U(1)_N$ gauge symmetry dates back to the 1970s, when it was still not clear  that Weinberg's $SU(2)_L \times U(1)_Y$ model was the right theory of electroweak interactions \cite{Lee:1977qs}. After the discovery of the $Z$ and $W$ gauge bosons,  since the electroweak gauge group is embedded into $SU(3)_L \times U(1)_N$, the so called 331 models \cite{Pisano:1992bxx,Frampton:1992wt}    became serious candidates to extend the SM and explain some issues for which it has no answer, such as the flavor problem   and a possible explanation for the  large splitting between the mass of the top quark and those of the remaining fermions. Several realizations of the 331 model have been proposed in the literature, which  predict new fermions, gauge bosons and scalar bosons,  so  their phenomenologies have  been considerably studied \cite{Ruiz-Alvarez:2012nvg,Kelso:2013nwa,RamirezBarreto:2011av,RamirezBarreto:2010vji,RamirezBarreto:2008wq,Phong:2013cfa,Dong:2014esa,Das:2020pai}.

The minimal 331 model \cite{Pisano:1992bxx,Frampton:1992wt} requires a very large scalar sector, which introduces  three scalar triplets to give masses to the new heavy gauge bosons and one scalar sextet to endow the leptons with  small masses. The complexity of this model has lead to the  appearance of alternative 331 models aimed to economize the scalar sector. In particular, the reduced 331 model  (RM331) \cite{Ferreira:2011hm} only requires two scalar triplets, thereby being considerably simpler than  the minimal version \cite{Cogollo:2008zc,Queiroz:2010rj}. In the RM331, the physical scalar states obtained after the symmetry breaking are two neutral scalar bosons only, with the lightest one being   identified with the SM Higgs boson \cite{Caetano:2013nya}, and a doubly charged one.  Unlike other 331 models, no singly charged scalar boson arises in the RM331 \cite{Dias:2013kma, Ferreira:2013nla,Vien:2013zra}. In the gauge sector, there are one new neutral gauge boson $Z'$, a new pair of  singly charged gauge bosons $V^\pm$, and a pair of doubly charged gauge bosons $U^{\pm\pm}$. Like other 331 models, the RM331 also predicts three new exotic quarks.
 The original RM331 is strongly disfavored by experimental data \cite{Huyen:2012uk}, though  it would  still be allowed as long as  left-handed quarks are introduced via a particular $SU(3)_L \times U(1)_N$ representation  \cite{Cogollo:2013mga,Kelso:2014qka}, which in fact would give rise to flavor changing neutral current (FCNC) effects.

The contributions to the electron and muon anomalous MDM have been already studied  in the RM331  \cite{Kelso:2013nwa} within  another 331 realization \cite{DeConto:2016ith}. As for the CMDM of quarks,  there is only a previous calculation in the context of an old version of the 331 model  \cite{Martinez:2007qf}, though  such a calculation is limited to the on-shell case. However, since the on-shell CMDM is infrarred divergent in the SM \cite{Hernandez-Juarez:2020drn}, a calculation of the off-shell CMDM is mandatory. To our knowledge there is  no calculation of the off-shell CMDM of quarks, let alone their off-shell CEDM, in 331 models.  Furthermore, in the model  studied in \cite{Martinez:2007qf}, the new contributions only arise in the gauge sector, whereas in the RM331 there are additional contributions from the neutral scalar bosons, which are absent in other 331 models. 

In this work we present a study on the contributions of the RM331 to the off-shell CMDM and CEDM of the top quark. Our manuscript is organized as follows. In Section II we present a brief description of the RM331, with the Feynman rules necessary for our calculation being presented in Appendix \ref{FeynmanRules}. The analytical calculation of the new contributions to the dipole form factors of the $\bar{t}tg$ vertex are presented in Sec. III; our results  in terms of Feynman parameter integrals and Passarino-Veltman scalar functions are presented in Appendix \ref{AnalyticalResults}. Section IV is devoted to a review of the current constraints on the parameter space of the model and the numerical analysis of the off-shell CMDM and CEDM of the top quark. Finally, in Sec. V the conclusions and outlook are presented.

\section{Brief outline of the RM331}\label{331model}
 We will describe briefly the main features of each sector of the RM331, focusing only on those details relevant to our calculation.

\subsection{Scalar and gauge boson eigenstates}
As far as the scalar sector is concerned, the scalar potential is given by
 \begin{equation}
V(\chi,\rho)=\mu_1^2 \rho^\dagger \rho +\mu_2^2\chi^\dagger\chi+\lambda_1\left(\rho^\dagger\rho\right)^2+\lambda_2\left(\chi^\dagger\chi \right)^2+\lambda_3\left(\rho^\dagger \rho \right)\left( \chi^\dagger\chi \right)+\lambda_4 \left(\rho^\dagger\chi \right)\left(\chi^\dagger\rho \right),
\end{equation}
where the scalar triplets transform as $\rho=\left(\rho^+,\rho^0,\rho^{++} \right)^T\sim (1,3,1)$ and $\chi=\left(\chi^-,\chi^{- -},\chi^0 \right)^T\sim(1,3,-1)$.  To induce the spontaneous symmetry breaking (SSB), the neutral scalar bosons $\rho^0$ and $\chi^0$ develop non-zero vacuum expectation values (VEVs)  under the shifting of the fields as
\begin{equation}
\rho^0,\chi^0\rightarrow \frac{1}{\sqrt{2}}\left(\upsilon_{\rho ,\chi}+R_{\rho ,\chi}+i I_{\rho ,\chi} \right),
\end{equation}
which leads to the following constraints
\begin{equation}
\begin{aligned}
\mu_1^2+\lambda_1\upsilon_\rho^2+\frac{\lambda_3\upsilon_\chi^2}{2}&=0,\nonumber\\
\mu_2^2+\lambda_2\upsilon_\chi^2+\frac{\lambda_3\upsilon_\rho^2}{2}&=0.
\end{aligned}
\end{equation}
The $SU(3)_C \times SU(3)_L \times U(1)_N$  breaks down into the SM gauge group following the pattern
\begin{equation}
 SU(3)_L \times U(1)_N\xrightarrow{\langle\chi^0\rangle}SU(2)_L \times U(1)_Y\xrightarrow{\langle\rho^0\rangle}U(1)_{\text{EM}},
\end{equation}
where $\upsilon_\rho$ can be identified with the SM Higgs VEV $\upsilon$.
The left-over of SSB are two neutral scalar bosons and a pair of doubly charged ones $h^{\pm\pm}$ as explained below.

The mass matrix of the neutral  scalar bosons in the $\left( R_\chi,R_\rho\right)$ basis is
\begin{equation}
\mathbf{m}_0^2=\frac{\upsilon_\chi^2}{2}\left(\begin{array}{cc}2\lambda_2 & \lambda_3 t \nonumber\\ \lambda_3 t & 2\lambda_1 t^2\end{array}\right),
\end{equation}
where $t=\upsilon_{\rho}/\upsilon_\chi$. After  diagonalization, the mass eigenstates in the limit $\upsilon_\chi\gg\upsilon_{\rho}$ are
\begin{equation}
h_1=c_\beta R_\rho -s_\beta R_\chi\text{,}\quad h_2=c_\beta R_\chi +s_\beta R_\rho,
\end{equation}
with masses 
\begin{align}
\label{mHiggs1}
m_{h_1}^2=\left( \lambda_1-\frac{\lambda_3^2}{4\lambda_2}\right)\upsilon_{\rho}^2,\\
\label{mHiggs2}
m^2_{h_2}= \lambda_2 \upsilon_\chi^2+\frac{\lambda_3^2}{4\lambda_2}\upsilon_{\rho},
\end{align}
 where $\lambda_1$, $\lambda_2>0$ and $c_\beta\equiv \cos\beta \approx 1-\lambda_3^2\upsilon^2_{\rho}/(8\lambda_2^2 \upsilon_\chi^2)$. The SM Higgs boson $h$ can be recovered in the $s_\beta\rightarrow 0$ limit, thus $h_1$ must be identified with the Higgs boson discovered at the LHC. Since $m_{h}\simeq 125$ GeV, from Eq. \eqref{mHiggs1} we obtain the relation $\lambda_1-\lambda_3^2/(4\lambda_2)\approx 0.26$ \cite{ParticleDataGroup:2020ssz}. In the case $\lambda_2$, $\lambda_2<1$, and $\lambda_3<\lambda_2$ we obtain  $m_{h_1}^2=\lambda_1 \upsilon_{\rho}^2$, which recovers the SM case and thus $\lambda_1\approx 0.26$.

In the gauge sector there are two new singly charged gauge bosons $V^\pm$, two doubly charged gauge bosons $U^{\pm\pm}$ and a neutral gauge bosons $Z'$. They acquire their masses as follows. The would-be Goldstone bosons $\chi^\pm$  are eaten by the singled charged gauge bosons, whereas a linear combination of the doubly charged would-be Goldstone bosons $\rho^{\pm\pm}$ and $\chi^{\pm\pm}$ are absorbed by the  doubly charged gauge boson $U^{\pm\pm}$. Also, the orthogonal combination of $\rho^{\pm\pm}$ and $\chi^{\pm\pm}$ gives rise to a physical doubly charged scalar boson pair $h^{\pm\pm}$. Finally, the would-be Goldstone boson $I_\chi$  becomes the longitudinal components of the $Z^\prime$ gauge boson. Thus, the masses of the new gauge bosons  at leading order at $\upsilon_{\chi}$ are \cite{Machado:2013jca}

\begin{align}
\label{massZprime}
m_{Z^\prime}^2&=\frac{g^2 c_W^2}{3(1-4s_W^2)}\upsilon_\chi^2,\\
\label{massV}
m^2_{V^\pm}&=\frac{g^2}{4}\upsilon^2_\chi,\\
m_{U^{\pm\pm}}^2&=\frac{g^2}{4}\left(\upsilon_\rho^2+\upsilon_{\chi}^2\right).
\end{align}

As far as the SM gauge bosons are concerned, the would-be Goldstone bosons $\rho^\pm$ and  $I_\rho$ endow with masses the  $Z$ and $W^\pm$ gauge bosons, respectively.

\subsection{Gauge and scalar boson couplings to the top quark}
The number of  new fermions  necessary to fill out the $SU(3)_L \times U(1)_N$ multiplets as well as their quantum numbers  depend on the particular  331 model version.
There are no new leptons in the RM331, but a new  quark is required for each quark triplet. They  transform as
\begin{align}
Q_{iL}=\begin{pmatrix}d_i\nonumber\\ -u_i\nonumber\\ J_i\end{pmatrix}_L\sim(3,3^\ast,-1/3),\quad i=1,2,\quad Q_{3L}=\begin{pmatrix}u_3\nonumber\\ d_3\nonumber\\ J_3\end{pmatrix}_L\sim(3,3,+2/3),
\end{align}
with the numbers between parentheses representing the field transformations under the $SU(3)_C \times SU(3)_L \times U(1)_N$  gauge group, whereas
$J_1$, $J_2$ and $J_3$ are the new exotic quarks with electric charges    $Q_{J_{1,2}}=-4/3e$ and $Q_{J_{3}}=5/3e$. Under this representation the theory is anomaly free \cite{Cogollo:2013mga}.

\subsubsection{Charged currents}
In the quark sector, the charged currents relevant for our calculation  are given by the following Lagrangian
\begin{align*}
\label{LagCC}
\mathcal{L}_q^{CC}&=\frac{g}{\sqrt{2}}\overline{u}_L V_{\text{CKM}}^q\gamma^\mu d_L W^+_\mu+\frac{g}{\sqrt{2}}\overline{J}_{3_L} \gamma^\mu\left(V^u_L\right)_{3a}u_{a_{L}}V^+_\mu+\frac{g}{\sqrt{2}}\overline{u}_{l_L}\left(V_L^{u\dagger}\right)_{li}\gamma^\mu J_{i_L}U^{++}_\mu+ {\rm H. c.},
\end{align*}
where the family index $a$ runs over 1, 2 and 3, whereas  $i$ and  $l$ run over 1 and 2.  Also $V_{\text{CKM}}^q=V_L^{u\dagger} V^d_L$ stands for the Cabibbo-Kobayashi-Maskawa matrix, with the mixing matrices $V^{u}_L$ ($V^{d}_L$) transforming the left-handed up (down) quarks flavor eigenstates into their mass eigenstates. It is assumed that the new quarks are given in their diagonal basis. Note that the doubly charged gauge boson $U^{\pm\pm}$ does not couples to the top quark.

\subsubsection{FCNC currents}

Since the  $Z^\prime$ gauge boson couplings to the  quarks  are non-universal, flavor changing neutral currents (FCNCs) are induced  at the tree level. The corresponding  Lagrangian for the up quark sector reads
\begin{equation}
\label{331ZFCNC}
\mathcal{L}^{FCNC}_{Z^\prime}=\frac{g}{2c_W\sqrt{3(1-4s_W^2)}}\left(\sum_{a=1}^3\left( \overline{u}^\prime_{aL}\gamma^\mu(1-2s_W^2)u^\prime_{aL}\right)+\overline{u}^\prime_{3L}\gamma^\mu(2s_W^2)u^\prime_{3L}\right)Z^\prime_{\mu},
\end{equation}
where the up quarks $u^\prime$ are in the flavor basis. It is evident that the above Lagrangian induces FCNC at the tree level after the rotation to the mass eigenstate basis.

On the other hand, the interactions between up quarks and the neutral scalar bosons arise from the lagrangian
\begin{equation}
\mathcal{L}_S=\sum_{i,j=1}^3\overline{u}^\prime_{iL}{\Gamma_1^u}_{ij} u_{Rj}^\prime h_1+\overline{u}^\prime_{iL}{\Gamma_2^u}_{ij}u_{jR}^\prime h_2
+ {\rm H. c.},
\end{equation}
where $u'$ is an up quark triplet $u'^T=(u',c',t')$ and
\begin{equation}
\boldsymbol{\Gamma}^u_1=\frac{c_\beta}{\upsilon_{\rho}}\mathbf{m}^u-\frac{s_\beta}{\upsilon_\chi}\left(\begin{array}{ccc}0 & 0 & 0 \nonumber\\0 & 0 & 0 \nonumber\\m_{31}^u & m_{32}^u & m_{33}^u\end{array}\right),
\end{equation}

\begin{equation}
\boldsymbol{\Gamma}^u_2=\frac{s_\beta}{\upsilon_{\rho}}\mathbf{m}^u+\frac{c_\beta}{\upsilon_\chi}\left(\begin{array}{ccc}0 & 0 & 0 \nonumber\\0 & 0 & 0 \nonumber\\m_{31}^u & m_{32}^u & m_{33}^u\end{array}\right),
\end{equation}
with $\mathbf{m}^{u}$ being the quark mass matrix in the flavor basis \cite{Cogollo:2013mga}. After rotating  to the mass eigenstate basis, only the terms proportional to $\mathbf{m}^{u}$ are diagonalized, whereas the remaining term gives rise to FCNC couplings, which  can be written as
\begin{equation}
\mathcal{L}^{FCNC}_{S}=\sum_{i,j=1}^3\left(-s_\beta \overline{u}_{iL}{\eta^u}_{ij}u_{jR}h_1
+c_\beta \overline{u}_{iL}{\eta^u}_{ij}u_{jR}h_2\right)+{\rm H. c.},
\end{equation}
where
\begin{equation}
{\boldsymbol{\eta}}^u=\mathbf{V}_L^u\left(\begin{array}{ccc}0 & 0 & 0 \nonumber\\0 & 0 & 0 \nonumber\\\frac{m_{31}^u}{\upsilon_\chi} & \frac{m_{32}^u}{\upsilon_\chi} & \frac{m_{33}^u}{\upsilon_\chi}\end{array}\right)\left(\mathbf{V}_R^u \right)^\dagger.
\end{equation}
Through  the parametrization given in \cite{Tatur:2008qt} for the $\mathbf{V}_{L,R}^{u,d}$ mixing matrices it is possible to obtain numerical values for the entries of the  ${\boldsymbol{\eta}}^{u,d}$ matrix. Under this framework $m_{31}^u=0$, $m_{32}^u=0$, and $m_{33}^u=m_t$.

\section{CMDM and CEDM of the top quark in the RM331}
\begin{figure}[H]
\begin{center}
\includegraphics[width=14cm]{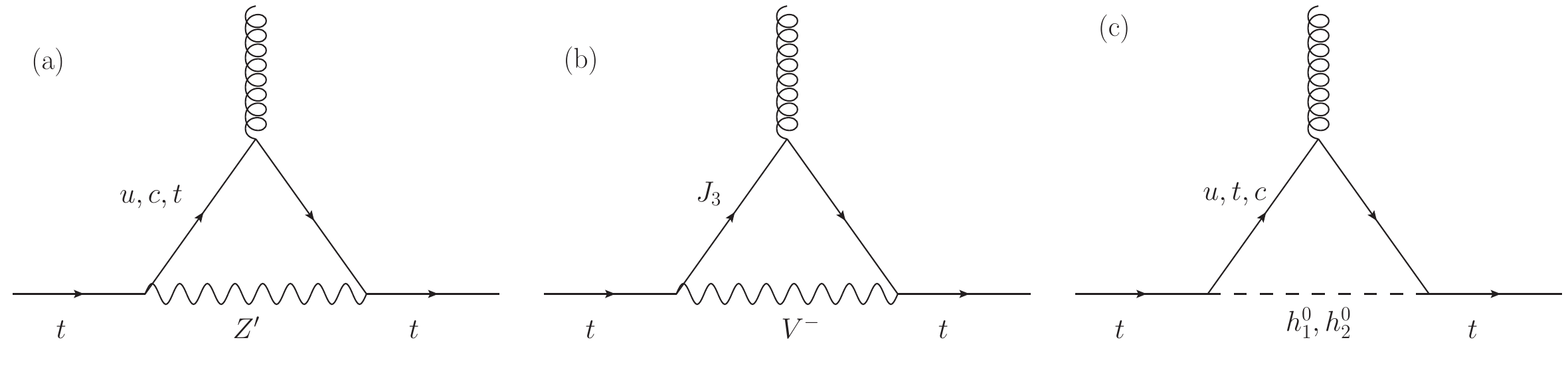}
\caption{New one-loop contributions of the RM331 to the CMDM and CEDM of the top quark  in the unitary gauge. In the conventional linear $R_\xi$ gauge there are additional Feynman diagrams where the gauge bosons are replaced by their associated Goldstone bosons. \label{FeynmanDiagrams}}
\end{center}
\end{figure}

Apart from the pure SM contributions, at the one-loop level there  are new contributions to the CMDM of the top quark arising in both the gauge and scalar sectors of the RM331.  The corresponding Feynman diagrams are depicted in Fig. \ref{FeynmanDiagrams}. In the gauge sector the new contributions arise  from  the neutral $Z'$ gauge boson, which are induced  by both diagonal and non-diagonal couplings. There are also a new contribution from the singly-charged gauge boson $V^\pm$, which is  accompanied by the new exotic quark $J_3$. As already noted, the doubly-charged gauge boson $U^{\pm\pm}$ does not couples to the top quark,  thus there is no contribution from this gauge boson  to the top quark CMDM and CEDM.  As for the scalar sector, there are new contributions   from the neutral scalar bosons $h_1$ and $h_2$,  which  in fact are the novel contributions from the RM331 as they are absent in other 331 model versions. The SM-like Higgs boson $h_1$ yields  new contributions arising from its FCNC couplings, which are induced  at the tree-level, but also from its diagonal coupling, which has a small deviation from its SM value. As for the new Higgs boson $h_2$, it also contributes via  both diagonal and non-diagonal couplings. We would like to point out that such scalar contributions are absent in the  331 model studied in Ref. \cite{Martinez:2007qf}, where the on-shell CMDM of the top quark was calculated. Even more, as long as complex  FCNC couplings are considered, there are  non-vanishing contributions to the CEDM. This class of contributions has also not been studied before in the context of 331 models.

We are interested in the off-shell CMDM and CEDM of the top quark.  Since off-shell Green functions  are not associated with an $S$-matrix element, they can be plagued by pathologies such as being gauge non-invariant, gauge dependent, ultraviolet divergent, etc. Along these lines, the pinch technique (PT) was meant to provide a systematic approach to construct well-behaved Green functions \cite{Binosi:2009qm}, out of which valid observable quantities can be extracted. It was later found that there is an equivalence at least at the one-loop level between the results found via the PT and those obtained through the background field method (BFM) via the Feynman gauge \cite{Denner:1994xt}. This provides a straightforward computational method to obtain gauge independent Green functions. It is thus necessary to verify whether  the RM331 contributions to the CMDM and CEDM of quarks are gauge independent for $q^2\ne 0$. Nevertheless, we note that from the Feynman diagrams of Fig. \ref{FeynmanDiagrams},  the gauge parameter $\xi$ only enters into the amplitudes  of the Feynman diagrams (a) and (b)
via the propagators of  the gauge bosons and their associated would-be Goldstone bosons. Those kind of diagrams have an amplitude that shares the same  structure to those mediated by the electroweak gauge bosons $Z$ and $W$ in the SM, which are known to yield a gauge independent contribution to the CMDM for an off-shell gluon when the contribution of their associated would-be Goldstone bosons are added up. See for instance Ref. \cite{Hernandez-Juarez:2020drn},  where we calculate the electroweak contribution to the CMDM of quarks in the conventional linear $R_\xi$ gauge and  verify that the gauge parameter $\xi$ drops out.
Furthermore, the dipole form factors cannot receive contributions from self-energy diagrams, which are required to cancel gauge dependent terms appearing in the monopolar terms via the PT approach. Thus both the CMDM and CEDM must be gauge independent for an off-shell gluon and thus  valid observable quantities.

Below we present the analytical results of our calculation in a model-independent way, out of which the results for the RM331 and other SM extensions would follow easily.  The  corresponding coupling constants for the RM331 are presented in Appendix \ref{FeynmanRules}.  For the loop integration we used  the Passarino-Veltman reduction method and for completeness our calculation was also performed  by  Feynman parametrization via the unitary gauge, which provides alternative expressions to cross-check the numerical results. The  Dirac algebra and the Passarino-Veltman reduction were done in Mathematica with the help of Feyncalc \cite{Shtabovenko:2016sxi} and Package-X \cite{Patel:2015tea}.

\subsection{New gauge boson  contributions}

We first consider the generic contribution of a new  gauge boson $V$ with the following  interaction to the quarks

\begin{equation}
\label{LagGauge}
{\cal L}^{Vqq'}=\frac{g}{c_W}\overline{q}\left(g_V^{Vqq'}-g_A^{Vqq'}\gamma^5\right)\gamma_\mu q' V^\mu+{\rm H.c.},
\end{equation}
where the coupling constants $g_{V,A}^{Vqq'}$ are taken in general as complex quantities. By hermicity they should obey  $g_{V,A}^{Vqq'}= g_{V,A}^{Vq'q*}$.

The above interaction gives rise to a new contribution to the quark CMDM and CEDM  via a Feynman diagram similar to that of  Fig. \ref{FeynmanDiagrams}(a). The corresponding contribution to the quark CMDM  can be written as
\begin{equation}
\label{aVq}
{\hat\mu}^{V}_q (q^2)=\frac{G_F m_W^2}{2\sqrt{2} \pi ^2 r_{V}^2 c_W^2}\sum_{q'} \left|g_V^{Vq q^\prime}\right|^2 \mathcal{V}_{qq'}^V(q^2)+
\left(\begin{array}{c}g_V^{Vq q^\prime}\to g_A^{Vq q^\prime}\\ m_q'\to -m_q'\end{array}\right),
%
\end{equation}
where we introduced the auxiliary variable $r_a=m_a/m_q$ and the $\mathcal{V}_{qq'}^V(q^2)$ function is presented in Appendix \ref{AnalyticalResults} in terms of Feynman parameter integrals and Passarino-Veltman scalar functions. The second term of the right-hand side stands for the first term with the indicated replacements.
As for the contribution to the quark CEDM, it can arise as long as there are flavor changing complex couplings and  is given by
   \begin{equation}
\label{dVq}
\begin{aligned}
{\hat d}^V_q(q^2)=\frac{G_F m_W^2}{\sqrt{2} \pi ^2r_{V}^2 c_W^2} \sum_{q'}\text{Im}\left(g_V^{Vq q^\prime} {g_A^{Vq q^\prime}}^\ast\right)\widetilde{\mathcal{D}}_{qq'}^V(q^2),
\end{aligned}
\end{equation}
where again the $\mathcal{D}_{qq'}^V(q^2)$ function is presented in Appendix \ref{AnalyticalResults}.

From Eqs. \eqref{aVq} y \eqref{dVq} we can obtain straightforwardly the contributions to the quark CMDM and CEDM of the neutral gauge boson $Z'$ and the singly charged gauge boson $V^\pm$ after  replacing  the coupling constants and the gauge boson masses. 

\subsection{New scalar boson contributions}

Following the same approach as above, we now present the generic contribution to the quark CMDM and CEDM arising from FCNC mediated by a new scalar boson $S$, which arise from the Feynman diagram of Fig \ref{FeynmanDiagrams}(c). We consider an interaction of the form

\begin{equation}
\label{LagScalar}
{\cal L}^{Sqq'}=-\frac{g}{2}\overline{q}\left(G_S^{Sqq'}+G_P^{Sqq'}\gamma^5\right) q' S+{\rm H.c.}
\end{equation}

The above scalar interaction leads to the following contribution to the quark CMDM
\begin{equation}
\label{Sq2CMDM}
{\hat\mu}^S_q(q^2)=-\frac{G_Fm_W^2}{8 \sqrt{2} \pi ^2}\sum_{q'}\left| G_P^{Sqq'}\right|^2 \mathcal{P}_{qq'}^S(q^2)+\left(\begin{array}{c}G_P^{Sqq'}\to G_S^{Sqq'}\\m_q'\to -m_q'\end{array}\right),
%
\end{equation}
whereas the corresponding contribution to the quark CEDM is given by
\begin{equation}
\label{Sq2CEDM}
{\hat d}^S_q(q^2)=\frac{G_Fm_W^2}{4 \sqrt{2} \pi
   ^2 } \sum_{q'}\text{Im}\left(G_S^{Sqq'} G_P^{Sqq'*} \right)\widetilde{\mathcal{D}}_{qq'}^S(q^2),
\end{equation}
where the $\mathcal{P}_{qq'}^S(q^2)$ and  $\widetilde{\mathcal{D}}_{qq'}^S(q^2)$ functions are presented in Appendix \ref{AnalyticalResults}.

From the above expression we can obtain the contribution of the new scalar Higgs boson of the RM331 as well as the contribution of the SM Higgs boson, which in the RM331 has tree-level FCNC couplings.

 \section{Numerical analysis and discussion}

We now turn to the numerical analysis. The coupling constants that enter into the  Feynman rules and are necessary to evaluate the CMDM and CEDM of the top quark [c.f. Eqs. \eqref{LagGauge} through \eqref{Sq2CEDM}] are presented in Tables \ref{FeynRulesBoson} and \ref{FeynRulesScalar} of Appendix \ref{FeynmanRules}. We note that these couplings depend on several free parameters, such as the mass parameter $m_{33}^u$, the VEV $\upsilon_\chi$,  the parameters of the scalar potential $\lambda_2$ and $\lambda_3$, as well as the entries of the matrices $\mathbf{V}_L^u$, $\mathbf{K}_L$ and ${\boldsymbol{\eta}}^u$. 
To obtain an estimate of the contributions of the RM331 to the CMDM and CEDM of the top quark we need to  discuss the most up-to-date constrains on these parameters from current experimental data.

 \subsection{Constraints on the parameter space}
\subsubsection{Heavy particle masses}

As already mentioned, the mass parameter $m_{33}^u$ can be identified with the top quark mass \cite{Cogollo:2013mga}, whereas  the VEV $\upsilon_\chi$ determines the masses of the heavy gauge bosons  and the heavy quark $J_3$. As for the mass of the new scalar boson $m_{h_2}$, it is determined by the parameters $\lambda_2$ and $\lambda_3$, along with the VEV $\upsilon_\chi$, which also determine  the mixing angle $s_\beta$. 

We will first discuss the current indirect constraints on the heavy neutral gauge boson masses.
From the muon $g-2$ discrepancy, the following constraint was obtained  $\upsilon_\chi\ge    2$ TeV \cite{Kelso:2014qka}, from which bounds on the heavy gauge boson masses follow. Nevertheless, there are also  indirect constrains obtained through the experimental data on $B^0-\overline{B}^0$ oscillations. The RM331 contribution to $\Delta m_B$ arises from FCNC couplings mediated by the $Z'$ gauge boson and the  $h_1$ and $h_2$ scalar bosons \cite{Cogollo:2013mga, Machado:2013jca}, then using the parametrization  of \cite{Tatur:2008qt},  the experimental limit on $\Delta m_B$ leads to the following bounds $m_{Z'}\gtrsim 3.3$ TeV, $m_{V^\pm}\gtrsim 0.33$ TeV and $m_{h_2}\gtrsim0.34$ TeV \cite{Cogollo:2013mga}. Similar limits have been imposed using de mass difference of the $K^0-\overline{K}^0$ and $D-\overline{D}^0$ systems \cite{Machado:2013jca}.
On the other hand, the current experimental bounds on the masses of new neutral and charged heavy gauge bosons from  collider searches  are model dependent \cite{ParticleDataGroup:2020ssz}. At the LHC, the ATLAS and CMS Collaborations have searched for an extra charged gauge boson  $W'$ at $\sqrt{s}=13$ TeV via the decay modes $W'\to \ell \nu_\ell$ \cite{ATLAS:2019lsy,CMS:2018hff} and $W'\to qq'$. The most stringent bounds  are obtained for a $W'$ gauge boson with SM couplings (sequential SM). The respective lower bounds on  $m_{W'}$ are 6.0 TeV (5.1 TeV) for the  $W'\to e \nu_e$   ($W'\to \mu \nu_\mu$) decay channel, whereas for the decay $W'\to q q'$ the corresponding bound is less stringent, of the order of $4$ TeV \cite{ATLAS:2019fgd,CMS:2019gwf}.
As far as an extra neutral gauge boson $Z'$ is concerned, the search at the LHC at $\sqrt {s}=13$ TeV via its decays into a lepton pair has been useful to impose the lower limit  $m_{Z'}\ge 4.5,5$ TeV for a $Z'$ gauge boson model arising in the sequential SM and in an $E_6$-motivated Gran Unification model \cite{ATLAS:2019erb,CMS:2021ctt}.  Along these lines,  it has been pointed out recently that the LHC might be able to constrain  the mass of the heavy $Z'$ boson up to the $5$ TeV level in several 331 models \cite{Barreto:2017xix,Long:2018fud,Long:2018dun}. Although these bounds are model dependent and relies on several assumptions,  if we consider the conservative value of 5 TeV for the gauge boson masses  we obtain a lower constraint on  $\upsilon_\chi$  of the order of 10 TeV. Thus, we will use this value in our analysis to be consistent with experimental constraints and limits from FCNC couplings.

As far as direct constraints on the mass of exotic quarks are concerned, the ATLAS and CMS Collaborations have used the $\sqrt{s}=13$ TeV data to search for vector-like quarks with electric charge of $5/3$ via its decay into a top quark and a $W$ gauge boson, with  the final state consisting of a single charged lepton (muon or electron), missing transverse momentum, and several jets. A mass exclusion limit up to $1.6$ TeV is obtained depending on the properties of the vector-like quark \cite{CMS:2018wpl,CMS:2018dcw,ATLAS:2018mpo}. We will thus use $m_{J_3}=2$ TeV to be consistent with the experimental bound.

\subsubsection{Mixing angle $s_\beta$ and parameters $\lambda_{2,3}$}

\begin{figure}[H]
\begin{center}
\includegraphics[height=6.5cm]{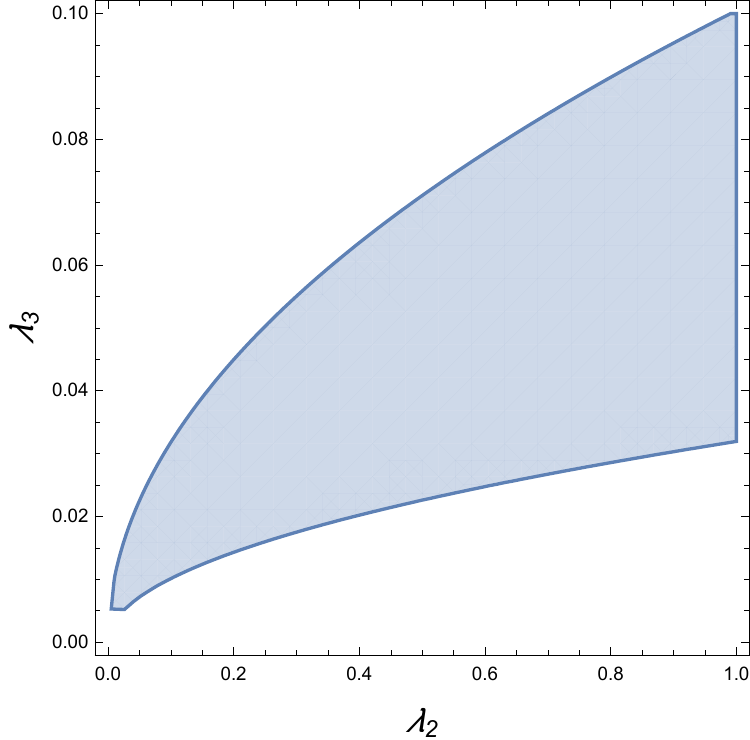}\hspace{0.5cm}\includegraphics[height=6.5cm]{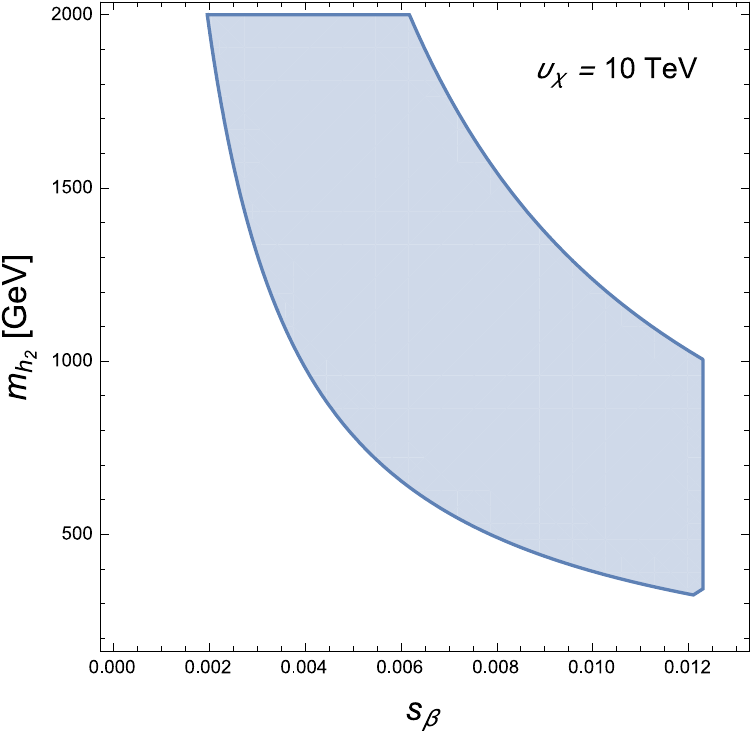}
\caption{Allowed areas in the planes $\lambda_3$ vs $\lambda_2$  and $s_\beta$ vs $m_{h_2}$  in agreement with the experimental error of the Higgs boson mass $m_h=125.10\pm 0.14$ GeV \cite{ParticleDataGroup:2020ssz} at 95\% C.L. We consider $\lambda_1\approx 0.26$ and $\lambda_3<\lambda_2<1$, which yield the SM limit.}
\label{Constrains}
\end{center}
\end{figure}

According to Eq. \eqref{mHiggs1} the mass of the SM-like Higgs boson receives new corrections through the $\lambda_2$ and $\lambda_3$ parameters. As discussed above, the SM case is recovered when $\lambda_1\approx0.26$ and $\lambda_3<\lambda_2<1$, thus the new corrections to $m_{h_1}$ must lie within the experimental error of the SM Higgs boson mass $m_h=125.10\pm 0.14$ GeV \cite{ParticleDataGroup:2020ssz}. This allows one to  constrain the $\lambda_2$ and $\lambda_3$ parameters, which in turn translates into constraints on $s_\beta$ and $m_{h_2}$ once the $\upsilon_\chi$ value is fixed.  Again we take a conservative approach and  only consider the experimental uncertainty in the Higgs boson mass, whereas theoretical uncertainties from higher order corrections are not taken into account.
We observe in Fig. \ref{Constrains} the allowed regions in the planes $\lambda_2$ vs $\lambda_3$ and $s_\beta$ vs $m_{h_2}$  consistent  with the experimental error of the Higgs boson mass at 95\% C.L. We note that for a given $\lambda_2$, $\lambda_3$ must be about one order of magnitude below.
In our calculation we use $\lambda_2=0.9$ and $\lambda_3=0.06$, though there is no great sensitivity of the top quark CMDM and CEDM to  mild changes in the values of these parameters. In addition, we find that  values ranging from $0.002$ to $0.013$ are allowed for $s_\beta$ provided that  $\upsilon_\chi\geq 10$ TeV and $m_{h_2}\gtrapprox 300$ GeV, which is consistent with recent searches for new neutral scalar bosons at the LHC \cite{ParticleDataGroup:2020ssz}.

\subsubsection{Mixing matrices}
As for the mixing matrices,  we can obtain the absolute values for the entries of the matrices $\mathbf{V}_L^u$, $\mathbf{K}_L$ and ${\boldsymbol{\eta}}^u$. The entries of the last matrix are given in terms of $\upsilon_\chi$, $s_\beta$ and the $m^q_{ij}$ matrix elements and their values are obtained following the parametrization used in \cite{Tatur:2008qt}. In general $\mathbf{K}_L$ and ${\boldsymbol{\eta}}^u$ are in terms of the entries of $\mathbf{V}_L^u$ and $\mathbf{V}_R^u$, the complex matrices that diagonalize the mass matrices of  up quarks. These matrices can be assumed to be triangular, then using the  experimental data on quark masses and the mixing angles it is possible to obtain values of their entries \cite{Tatur:2006aw}.    It is also assumed that the only non-negligible  mixing is that arising between the third and second fermion families.  Furthermore, since the $CP$ violation phases are expected to  be very small, we take a conservative approach and assume  complex phases of the order of $10^{-3}$.

We present in  Table \ref{Parameters} a summary of the numerical values we will use in our numerical evaluation.

\begin{table}[H]
  \centering
  \caption{Values of the parameters used in our evaluation of the CMDM  and CEDM of the top quark in the RM331. For the entries of the matrices $\mathbf{V}_L^u$, $\mathbf{K}_L$ and ${\boldsymbol{\eta}}^u$  we use the values obtained in  \cite{Cogollo:2013mga} using the parametrization of \cite{Tatur:2008qt}, where the mass parameter $m_{33}^u$ is identified with the top quark mass. We use $\lambda_2$ and $\lambda_3$ values allowed by the experimental error in the Higgs boson mass and also assume that the only non-negligible  mixing is that arising between the third and second fermion families.  }\label{Parameters}
 \begin{tabular}{cc}
  \\\hline\hline  Parameter & Value
  \\\hline\hline  $\left|(K_L)_{tc}\right|$ & $6.4\times10^{-4}$
    \\$\left|V^u_{33}\right|$ &1
    \\ $\left|{\eta}^u_{tc}\right|$ & $6.4\times 10^{-4}$
     \\$\left|{\eta}^u_{ct}\right|$ & $4.62\times 10^{-6}$
      \\  $m_{33}^u$ & $m_t$
  \\    $\upsilon_\chi$&10 TeV
  \\$s_\beta$&$10^{-2}$
  \\ $m_{h_2}$&300 GeV
 \\$\phi_{{{\eta}}^u_{tc}}$, $\phi_{{{\eta}}^u_{ct}}$ & $10^{-3}$
  \\\hline\hline
  \end{tabular}
  \end{table}

 \subsection{Top quark CMDM}

As already mentioned, in the RM331  there are new contributions to the off-shell top quark CMDM $\mu_t(q^2)$ arising from the heavy gauge bosons $Z'$ and $V^\pm$ as well as the neutral scalar bosons $h_1$ and $h_2$. Below we will use the notation $A_{BC}$ for the  contribution of particle $A$  due to the  $ABC$ coupling. Thus, for instance $Z'_{tc}$ will denote the contribution of the loop with the $Z'$ gauge boson due to the $Z't c$ coupling. Since we would like to assess the magnitude of the new physics  contributions  to $\hat \mu_t(q^2)$,  we  will extract from our calculation the pure SM contributions. Thus, apart from the contribution due to the tree-level FCNCs of the SM-like Higgs boson $h_1$, we only consider the  contribution arising from the small deviation of the diagonal coupling $h_1tt
$ from the SM $htt
$ coupling. This contribution will be denoted by $\delta{h_1}_{tt}$.

\begin{figure}[H]
\begin{center}
{\includegraphics[height=4cm]{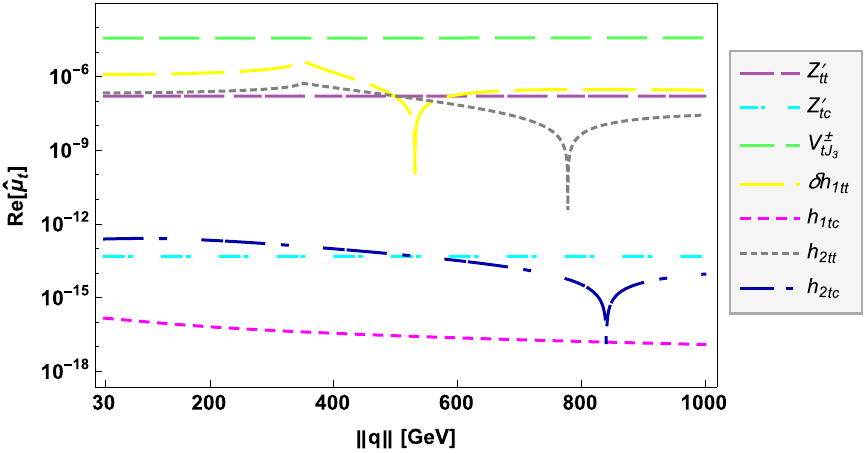}}
{\includegraphics[height=4cm]{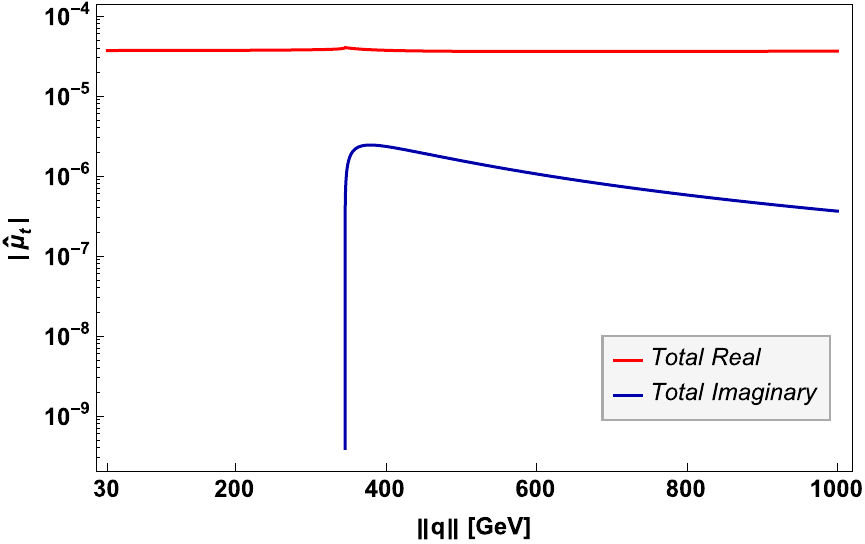}}
\caption{Real part of the partial contributions of the RM331 to the top quark CMDM (left plot) as a function of $\|q\|\equiv\sqrt{|q^2|}$ for the parameter values of Table \ref{Parameters}. The  real and imaginary parts of the total contribution are shown in the right plot.}
\label{CMDM_q}
\end{center}
\end{figure}

We will examine the behavior of the CMDM of the top quark as a function of $\|q\|\equiv \sqrt{|q^2|}$, where $q$ is the gluon four-momentum. In  the left plot of Fig. \ref{CMDM_q} we show  the real part of the partial contributions  to $\hat\mu_t(q^2)$ as a function of  $\|q\|$  for the parameter values of Table \ref{Parameters}, whereas the real and imaginary parts of the total contribution are shown in the right plot.  In general there is little dependence of  ${\rm Re}\left[\hat\mu_t(q^2)\right]$ on $\|q\|$,  except for  the $\delta{h_1}_{tt}$, $h_{2tt}$ and $h_{2tc}$ contributions, which have a change sign. We also note that the  $V^\pm_{tJ_3}$ contribution is the largest one, whereas the  remaining contributions are negligible, with the ${h_1}_{tc}$ contribution being the smallest one. Thus the curve for the real part of the total contribution  seems to overlap with that of the $V^\pm_{tJ_3}$ contribution, though the former shows a small peak at $\|q\|\simeq 2 m_t$. This  can be explained by the peak appearing in the  $\delta{h_1}_{tt}$ contribution, which can be as large as the ${V^\pm}_{tJ_3}$   contribution for $\|q\|\simeq 2 m_t$. We conclude that  $\hat \mu_t(q^2)$ can have  a real part  of the order of  $10^{-5}$.  

As far as the imaginary parts of the partial contributions to $\hat\mu_t(q^2)$,  they are several orders of magnitude smaller than the corresponding real parts. As observed in the right plot of Fig. \ref{CMDM_q}, the imaginary part of the total contribution is negligible for $\|q\|\leqslant 2 m_t$, but increases up to about $10^{-6}$ around $\|q\|=400$ GeV, where  it starts to decrease up to  one order of magnitude as $\|q\|$ increases up to 1 TeV.

\begin{figure}[H]
\begin{center}
\includegraphics[height=4.5cm]{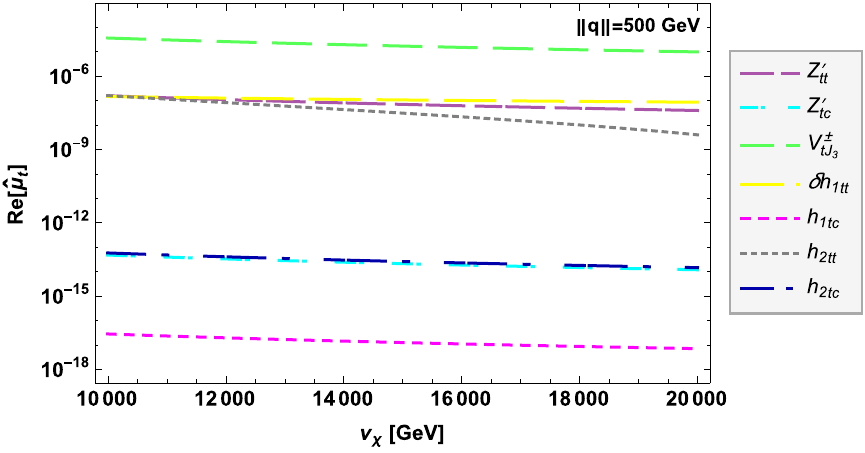}
\includegraphics[height=4.5cm]{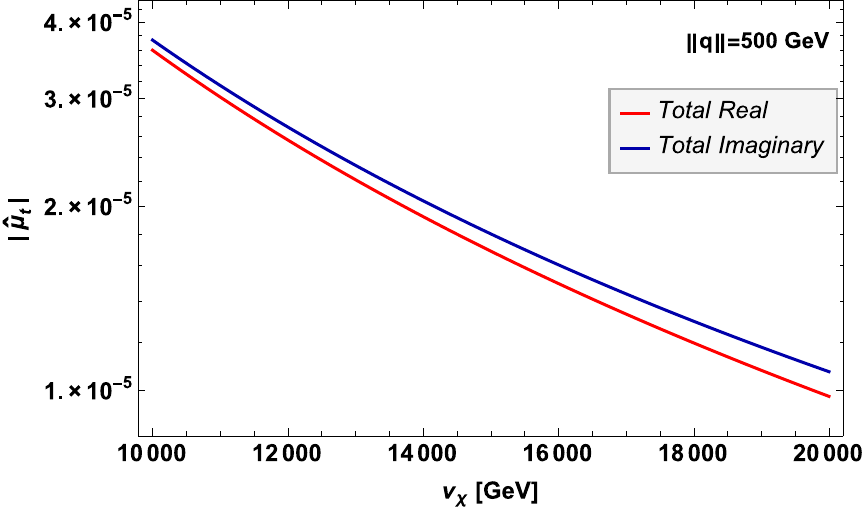}
\caption{The shames as in Fig. \ref{CMDM_q} but for the contributions of the RM331 to the top quark CMDM as functions of $\upsilon_\chi$ for $\|q\|$=500 GeV. For the remaining parameters we use the values of Table \ref{Parameters}.}
\label{CMDM_upsilon}
\end{center}
\end{figure}

Analogue plots to those of Fig. \ref{CMDM_q}, but now for the behavior of $\hat\mu_t(q^2)$ as a function of $\upsilon_\chi$ for  $\|q\|=500$ GeV and the parameter values of Table \ref{Parameters}, are shown  in Fig. \ref{CMDM_upsilon}.  In this case we observe that the real parts of the partial contributions to $\hat\mu_t(q^2)$ show a variation of about one order of magnitude when $\upsilon_\chi$ increases from 10 TeV to 20 TeV. As already noted, the $V^\pm_{tJ_3}$ contribution yields the bulk of the  total contribution to $\hat \mu_t$, whose imaginary part is slightly larger than its real part. Therefore both real and imaginary contributions of the RM331 to the top quark CMDM can be as large as  $10^{-5}$.

In summary,  for $\upsilon_\chi\geqslant 10$ TeV the real part of the  the RM331 new contribution to $\hat\mu_t(q^2)$ would be three orders of magnitude smaller than  the real part of the SM electroweak contribution \cite{Hernandez-Juarez:2020drn}, whereas its imaginary part  can be as large than its real part.  In general there is no appreciable variation in the magnitude of $\hat\mu_t$ for mild changes in the parameters of Table \ref{Parameters}. Although $\hat\mu_t(q^2)$ can be of similar size than the SM electroweak prediction for $\upsilon_\chi\le 10$  TeV, such values are disfavored by the current constrains on the heavy gauge bosons masses. Finally, we  note that  the RM331 can give a contribution  larger than the ones predicted by other extension models where a new neutral $Z$ gauge boson is predicted \cite{Aranda:2018zis}. The real and imaginary parts of the top quark CMDM are of order $10^{-6}-10^{-7}$ and $10^{-10}-10^{-11}$ respectively in such models.

\subsection{Top quark CEDM}

A potential new source of $CP$ violation can arise in the RM331 through the FCNC couplings mediated by the neutral scalar bosons, which are proportional to the entries of the non-symmetric complex mixing matrix ${\boldsymbol{\eta}}^u$  \cite{Cogollo:2013mga}, thereby allowing the presence of a non-zero CEDM, which is  absent in other 331 models. Thus, it is a novel prediction of the RM331. 

\begin{figure}[H]
\begin{center}
{\includegraphics[width=14cm]{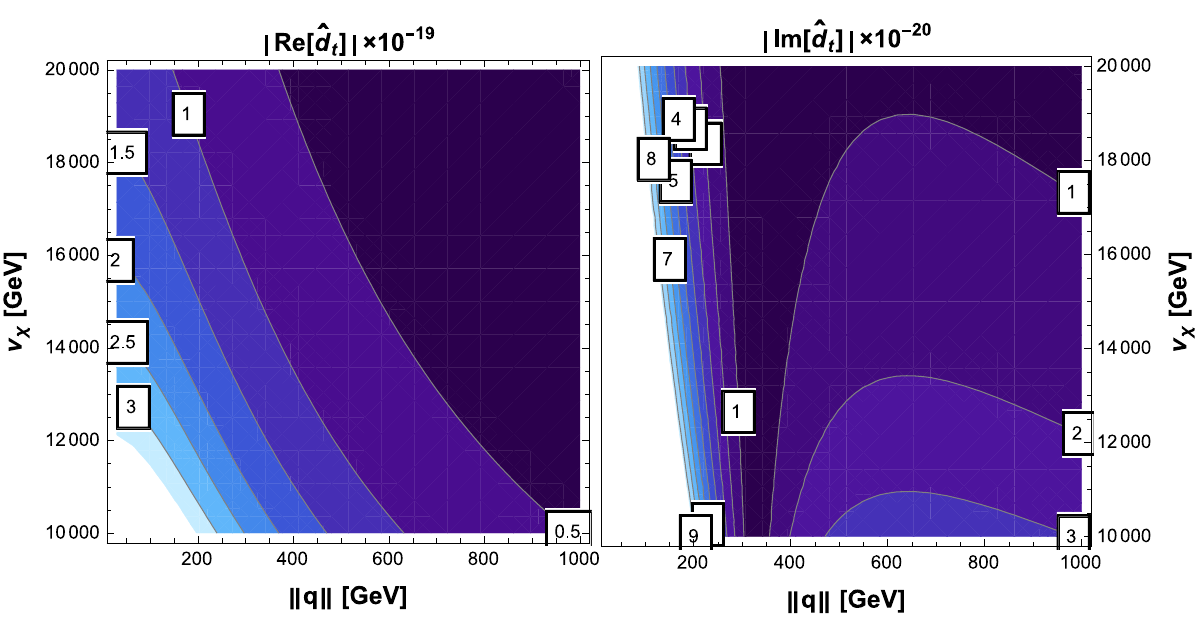}}
\caption{Real part (left plot) and imaginary part (right plot) of the total contribution to the CEDM of the top quark in the RM331 in the plane $\upsilon_\chi$ vs $\|q\|$. We use the parameter values of Table \ref{Parameters}.}
\label{ElectricPlot1}
\end{center}
\end{figure}

There are only two partial contributions to the top quark CEDM in the RM331, thus we only analyze the behavior of the total contribution. We
 show in Fig. \ref{ElectricPlot1} the contour lines of the real part (left plot) and the imaginary part (right plot)  of $d_t(q^2)$ in the $\upsilon_\chi$ vs $\|q\|$ plane for the parameter values of Table \ref{Parameters}.  We have found that  the new scalar  boson $h_2$ yields the dominant  contribution to $d_t(q^2)$, whose real (imaginary) part can be as large as $10^{-19}$ ($10^{-20}$), whereas the contribution from the $h_1$ scalar boson is  three or more orders of magnitude below.  We also observe that the real part of $d_t(q^2)$ decreases as  $\upsilon_\chi$ and $\|q\|$ increase, while the imaginary part remains almost constant. For $\|q\| \geqslant 600$ GeV, the RM331 contribution to the CEDM of the top quark is expected to be below the $10^{-20}$  level, which seems to be much smaller than the values predicted in other extension models \cite{Aranda:2018zis}, where the real and imaginary parts are of order $10^{-7}-10^{-8}$ and $10^{-12}-10^{-13}$ respectively.   In the range 2 TeV$\leqslant\upsilon_\chi\lesssim$ 10 TeV our results for $d_t(q^2)$ are enhanced by one order of magnitude, but as already noted, this interval is disfavored by current constraints.

For comparison, a compilation of the predictions of several extension models  of the top quark CMDM and CEDM for $q^2=0$  is presented in Table \ref{CMDMPred}. We  would like to stress that to our knowledge there is no previous estimate of the top quark CEDM in 331 models. We also note that though these values seem to be much  larger than the results  obtained for $q^2\neq0$ in the RM331, the dipole form factors are expected to decrease as $q^2$ increases. Such a behavior is indeed  observed in the SM case \cite{Hernandez-Juarez:2020drn}, where the magnitude of $\hat{a}_t$ decreases as $\|q\|$ increases.

\begin{table}[htb!]
\begin{center}
\centering
\caption{Predictions of the CMDM and CEDM of the top quark in several extension models at $q^2=0$.  }
\begin{tabular}{cccc}
\hline
\hline
 Model & $\hat{a}_t$ && $\hat{d}_t$    \\
\hline
\hline
SM & $10^{-2}$ \cite{Hernandez-Juarez:2020drn,Aranda:2020tox}&&   \\
\hline
THDMs & $10^{-3}$--$10^{-1}$ \cite{Gaitan:2015aia} &&$10^{-5}$ \cite{Iltan:2001vg, Gaitan:2015aia}\\
\hline
4GTHDM &{$10^{-2}$--$10^{-1}$ } \cite{Hernandez-Juarez:2018uow}&&{$10^{-5}$--$10^{-4}$} \cite{Hernandez-Juarez:2018uow}
\\
\hline
331 & $10^{-5}$ \cite{Martinez:2007qf} &&\\
\hline
Technicolor & $10^{-2}$ \cite{Martinez:2007qf}&& \\
\hline
Extra dimensions & $10^{-3}$ \cite{Martinez:2007qf}&& \\
\hline
Little Higgs model & $10^{-6}$ \cite{Ding:2008nh}&& \\
\hline
MSSM &$10^{-1}$ \cite{Aboubrahim:2015zpa}  && $10^{-5}-10^{-4}$ \cite{Atwood:2000tu} \\
\hline
Unparticle model &$ 10^{-2}$ \cite{Martinez:2008hm}  &&  \\
\hline
 Vector-like multiplets &  && $10^{-4}$ \cite{Ibrahim:2011im} \\
\hline
\hline
\end{tabular}
\label{CMDMPred}
\end{center}
\end{table}

\section{Conclusions}
We have presented a calculation of the one-loop contributions to the CMDM and CEDM, $\hat\mu_t(q^2)$ and $\hat d_t(q^2)$, of the top quark in the framework of the RM331, which is an economic version of the so-called 331 models with a scalar sector comprised by two scalar triplets only. We have considered the general case of an off-shell gluon as it has been pointed out before that the QCD contribution to $\hat\mu_t(q^2)$ is infrared divergent and the CMDM has no physical meaning for $q^2=0$. We argue that the results are gauge independent for $q^2\ne 0$  and represent valid observable quantities since the structure of the gauge boson contributions are analogue to those arising in the SM. To our knowledge, no previous calculations of the off-shell CMDM and CEDM of the top quark have been presented before in the context of 331 models.

Apart from the usual SM contributions, in the RM331, the CMDM  of the top quark receives new contributions from two new  heavy gauge bosons $Z'$ and $V^\pm$ as well as  one new neutral scalar boson $h_2$, along with a new contribution from the neutral scalar boson $h_1$, which must be identified with the 125 GeV scalar boson detected at the LHC. This model also predicts tree-level FCNCs mediated by the $Z'$ gauge boson and the two neutral scalar bosons $h_1$ and $h_2$, which at the one-loop level can also give rise to a non-vanishing CEDM provided that there is a $CP$-violating phase. The analytical results are presented in terms of both Feynman parameter integrals and Passarino-Veltman scalar functions, which are useful to cross-check the numerical results.

We present an analysis of the region of the parameter space of the model  consistent with experimental data and evaluate the CMDM and CEDM of the top quark for parameter values still allowed. It is found that the new one-loop contributions of the RM331 to the real (imaginary) part of $\hat \mu_t(q^2)$ are of  order of  $10^{-5}$ ($10^{-6}$), which are larger than the predictions of other SM extensions \cite{Aranda:2018zis}, with the dominant contribution arising from the $V^\pm$ gauge boson, whereas the remaining contributions are considerably smaller. It is also found that there is little dependence of $\mu_t(q^2)$ on $\|q\|$ in the  30-1000 GeV interval for a mass  $m_{V}$ of the order of a few hundreds of GeV. As far as the CEDM of the top quark is concerned, it is mainly induced by the loop with $h_2$ exchange and can reach values of the order of $10^{-19}$ for realistic values of the $CP$-violating phases. Such a contribution is smaller than the ones predicted by other SM extensions \cite{Aranda:2018zis}.

\acknowledgments{We acknowledge support from Consejo Nacional de Ciencia y Tecnolog\'ia and Sistema Nacional de Investigadores. Partial support from Vicerrector\'ia de Investigaci\'on y Estudios de Posgrado de la Ben\'emerita Universidad Aut\'onoma de Puebla is also acknowledged. }

\appendix
\section{Feynman rules}
\label{FeynmanRules}
We now present in Tables \ref{FeynRulesBoson} and \ref{FeynRulesScalar} the coupling constants that enter into the  Feynman rules  \cite{Cogollo:2013mga,Machado:2013jca,Kelso:2013nwa} that follow from Eqs. \eqref{LagGauge} and \eqref{LagScalar} and are necessary for the evaluation of the CMDM and CEDM of the top quark in the RM331.

\begin{table}[H]
  \centering
  \caption{Coupling constants for the interactions between gauge bosons and quarks in the RM331. We follow the notation of Lagrangian \eqref{LagGauge}. Here  $(K_L)_{tq}$ are entries of the complex mixing matrix $\mathbf{K}_{L}$,  where the subscript $q$ runs over  $u$ and $c$. This matrix is given in terms of the unitary complex matrix $\mathbf{V}^u_L$ that diagonalizes the mass matrix of up quarks, and can be written as $(K_L)_{tq}=(V^u_L)^\ast_{tq}(V^u_L)_{qt}$. Here $h_W=1-4s_W^2$.}\label{FeynRulesBoson}
 \begin{tabular}{ccc}\hline\hline Coupling &$g_V^{Vqq^\prime}$ & $g_A^{Vqq^\prime}$
  \\\hline\hline  $Z^\prime \overline{t} t$ & $\frac{1-2s_W^2}{2\sqrt{12 h_W}} $& $\frac{1-2s_W^2}{2\sqrt{12 h_W}}$
 \\  $Z\overline{t}q$&$\frac{s_W^2}{\sqrt{12 h_W}}\left(K_L\right)_{tq}$ & $\frac{s_W^2}{\sqrt{12 h_W}}\left(K_L\right)_{tq}$ \\
  $V^- \overline{t}J_3$&$\sqrt{2}c_W (V^u_L)_{33}$&$\sqrt{2}c_W (V^u_L)_{33}$
  \\\hline\hline
  \end{tabular}
  \end{table}

\begin{table}[H]
  \centering
  \caption{Coupling constants for the interactions between scalar bosons and quarks necessary for the evaluation of  the one-loop contributions to the CMDM and CEDM in the RM331. We follow the notation of Lagrangian \eqref{LagScalar}. Here $({\eta}^u)_{tq}$ are entries of the complex mixing matrix ${\boldsymbol{\eta}}^u$, where the subscript $q$  runs over $u$ and $c$. This matrix is given in terms of the unitary complex matrices $\mathbf{V}^u_L$ and $\mathbf{V}^d_L$ that diagonalize the mass matrix of up quarks, and can be written as $({\eta}^u)_{tq}=(V_L^u)_{q q}(V_R^u)^\ast_{t q}$ and $({\eta}^u)_{qt}^\ast=(V_L^u)^\ast_{t q}(V_R^u)_{q q}$ since the matrix ${\boldsymbol{\eta}}^u$ is not symmetric. }\label{FeynRulesScalar}
 \begin{tabular}{ccc}
  \\\hline\hline  & $G_S^{Sqq^\prime}$ & $G_P^{Sqq^\prime}$
  \\\hline\hline $h_1\overline{t}t$ & $\frac{m_t}{m_W} \left(c_\beta-\frac{\upsilon_{\rho}}{\upsilon_\chi}s_\beta\right)$& -\\
   $h_1\overline{t}q$ &$-\frac{s_\beta \upsilon_{\rho} m_{33}}{\upsilon_\chi\,m_W} \left(({\eta}^u)_{tq}+ ({\eta}^u)^\ast_{qt}\right)$ & $-\frac{s_\beta \upsilon_{\rho} m_{33}}{\upsilon_\chi\,m_W} \left(({\eta}^u)_{tq}- ({\eta}^u)^\ast_{qt}\right)$
  \\ $h_2\overline{t}t$&$\frac{m_t}{m_W} \left(s_\beta-\frac{\upsilon_{\rho}}{\upsilon_\chi}c_\beta\right)$&-
  \\ $h_2\overline{t}q$ &$\frac{c_\beta \upsilon_{\rho} m_{33}}{\upsilon_\chi\,m_W} \left(({\eta}^u)_{tq}+ ({\eta}^u)^\ast_{qt}\right)$ & $\frac{c_\beta \upsilon_{\rho} m_{33}}{\upsilon_\chi\,m_W} \left(({\eta}^u)_{tq}- ({\eta}^u)^\ast_{qt}\right)$
  \\\hline\hline
  \end{tabular}
  \end{table}

\section{Analytical results for the loop integrals}
\label{AnalyticalResults}
In this appendix we present the loop integrals appearing in Eqs. \eqref{aVq}, \eqref{dVq}, \eqref{Sq2CMDM}, and \eqref{Sq2CEDM} in terms of Feynman parameter integrals and Passarino-Veltman scalar functions both for non-zero and zero $q^2$. We have verified that all the ultraviolet divergences cancel out. Furthermore,  contrary to the QCD contribution, all the contribution of the RM331 are finite for $q^2=0$.

\subsection{Feynman parameter integrals}\label{FeynIntegrals}
The $\mathcal{V}_{qq'}^V(q^2)$  function of Eq. \eqref{aVq} can be written as

\begin{align}
\mathcal{V}_{qq'}^V(q^2)=&\int_0^1\int_0^{1-u} \frac{dudv}{\Delta_V}\Big[2  (u-1)^2 u+(1-r_{q^\prime}) \left(2 {\hat q}^2 u v (u+v-1)+(3 u-1) \Delta_V \log
   \left({\Delta_V}\right)\right)\nonumber\\
   &- r_{q^\prime} (u-1)^2 (2 u-1)-\left(2 r_{q^\prime}^2
   (u-1)^2+r_{V}^2 u (u+3)\right)+r_{q^\prime} \big(r_{q^\prime}^2 (u-1)^2-r_{V}^2
   (u-5) u\big)\Big],
\end{align}
where
$\Delta_V =u \left((u-1)+r_{V}^2\right)-r_{q^\prime}^2 (u-1)+{\hat q}^2 v
   (u+v-1)$, with  ${\hat q}^2=q^2/m_q^2$ and  $r_a=m_a/m_q$.

For $q^2=0$ we obtain
\begin{align}
\mathcal{V}_{qq'}^V(0)=&\int^1_0 \frac{udu }{r_{V}^2 (u-1)-u \left(
   (u-1)+r_{q^\prime}^2\right)} \Big[u^2 \left((1-r_{q^\prime})^2+2 r_{V}^2\right)-u \left(2 r_{V}^2
   (3 -2 r_{q^\prime})+(1+r_{q^\prime}) (1-r_{q^\prime})^2\right)\nonumber\\
   &+4 r_{V}^2
   (1-r_{q^\prime})\Big].
\end{align}

As far as the $\widetilde{\mathcal{D}}_{qq'}^V(q^2)$ function of Eq. \eqref{dVq} is concerned, it is given by
\begin{align}
\widetilde{\mathcal{D}}_{qq'}^V(q^2)=& r_{q^\prime}\int_0^1\int_0^{1-u}\frac{dudv}{\Delta_V} \Big[(3 u-1) \Delta_V \log
   \left({\Delta_V}\right)+(2 u+1) (u-1)^2-r_{q^\prime}^2 (u-1)^2\nonumber\\
   &+u \left(r_{V}^2
   (u-5)+2 {\hat q}^2 v (u+v-1)\right)\Big],
\end{align}
which leads to
\begin{align}
\widetilde{\mathcal{D}}_{qq'}^V(0)=r_{q^\prime}\int^1_0 \frac{ u\left(u (1-r_{q^\prime}^2)+4 r_{V}^2 (u-1)\right)}{u \left(
   (u-1)+r_{q^\prime}^2\right)-r_{V}^2 (u-1)} du.
\end{align}

The $\mathcal{P}_{qq'}^S(q^2)$ function  of Eq. \eqref{Sq2CMDM} is
\begin{equation}
\mathcal{P}_{qq'}^S(q^2)=\int_0^1\int_0^{1-u}\frac{ (u-1)  (u-r_{q^\prime})}{ u \left( (u-1)+r_{S}^2\right)-r_{q^\prime}^2 (u-1)+{\hat q}^2 v (u+v-1)}dudv,
\end{equation}
which for $q^2=$ simplifies to
\begin{equation}
\mathcal{P}_{qq'}^S(0)=\int_0^1\frac{ u^2  ( (1-u)-r_{q^\prime})}{u \left(
   (u-1)+r_{q^\prime}^2\right)-r_{S}^2 (u-1)}du.
\end{equation}

Finally, the loop function of Eq. \eqref{Sq2CEDM} reads
\begin{equation}
\widetilde{\mathcal{D}}_{qq'}^S(q^2)=\int_0^1\int_0^{1-u}\frac{ r_{q^\prime} (u-1)}{u \left( (u-1)+r_{S}^2\right)-r_{q^\prime}^2 (u-1)+{\hat q}^2 v
   (u+v-1)}dudv,
\end{equation}
which yields
\begin{equation}
\widetilde{\mathcal{D}}_{qq'}^S(0)=\int^1_0\frac{ r_{q^\prime}(1-u)^2 }{ (1-u) \left(r_{q^\prime}^2- u\right)+r_{S}^2
   u}du.
\end{equation}

\subsection{Passarino-Veltman results}
\label{PV}
We now present the results for the  loop functions in terms of Passarino-Veltman scalar functions, which can be numerically evaluated  by either LoopTools  \cite{Hahn:1998yk} or Collier \cite{Denner:2016kdg}, which allows one to cross-check the results. We introduce the following notation for the two- and three-point scalar functions in the customary notation used in the literature:

\begin{align}
\label{scalarfunctions}
B_{a}&=B_0(0,m_a^2,m_a^2),\\
B_{q^\prime b}&= B_0(m_q^2,m_{q^\prime}^2,m_b^2), \\
B_{\hat{q}q^\prime}&=B_0(\hat{q}^2,m_{q^\prime}^2,m_{q^\prime}^2),\\
C_{a}&=m_q^2C_0(m_q^2,m_q^2,q^2,m_{q^\prime}^2,m_a^2,m_{q^\prime}^2).
\end{align}
for $a=V,S,q^\prime$ and $b=V,S$. We also define $\delta_b=1-r_b$ and $\chi_b=1+r_b$.

For non-zero $q^2$, the loop functions of Eqs. \eqref{aVq} and \eqref{dVq} are given by
\begin{align}
\mathcal{V}_{qq'}^{V}(q^2)&=\frac{1}{\left(\hat{q}^2-4\right)^2}\Big[
\left(\hat{q}^2-4\right) \left(r_{q^\prime}^2-r_V^2+1\right) \left(\delta _{q^\prime}^2+2 r_V^2\right)+
\left(\hat{q}^2-4\right) \left(\delta _{q^\prime}^2+2 r_V^2\right)\left(r_{q^\prime}^2 B_{q^\prime}- r_V^2B_V\right)
\nonumber\\& - \Big(\delta _{q^\prime}^2 \chi _{q^\prime} \left(\left(\hat{q}^2-10\right) r_{q^\prime}+\hat{q}^2+2\right)+r_V^2 \left(\hat{q}^2 \left(r_{q^\prime}-3\right){}^2-2 r_{q^\prime} \left(5 r_{q^\prime}-6\right)-18\right)-2 \left(\hat{q}^2-10\right) r_V^4\Big)B_{q^\prime V}\nonumber\\& + \Big(\delta _{q^\prime}^2 \left(2 \hat{q}^2 r_{q^\prime}-2 r_{q^\prime} \left(3 r_{q^\prime}+4\right)+\hat{q}^2+2\right)-2 r_V^2 \left(\hat{q}^2 \left(4 r_{q^\prime}-5\right)+r_{q^\prime} \left(3 r_{q^\prime}-10\right)+11\right)+12 r_V^4\Big)B_{\hat{q}{q^\prime}}
\nonumber\\&+2
\Big(\delta _{q^\prime}^3 \chi _{q^\prime}^2 \left(3 r_{q^\prime}-\hat{q}^2+1\right)+\delta _{q^\prime} r_V^2 \left(\left(5 \hat{q}^2-8\right) r_{q^\prime}^2-\left(\hat{q}^2-4\right) r_{q^\prime}-2 \left(\left(\hat{q}^2-4\right) \hat{q}^2+6\right)\right)\nonumber\\&
-r_V^4 \left(4 \hat{q}^2 \left(r_{q^\prime}-2\right)-\left(10-9 r_{q^\prime}\right) r_{q^\prime}-17\right)+6 r_V^6\Big)C_{q^\prime V}
\Big],
\end{align}
and
\begin{align}
\widetilde{\mathcal{D}}_{qq'}^{V}(q^2)&=\frac{r_{q^\prime}}{\hat{q}^2-4}\Big[
\Big(r_{q^\prime}^2-4 r_V^2-1\Big)\left(B_{q^\prime V}-B_{\hat{q}{q^\prime}}\right) + \Big(r_V^2 \left(2 \hat{q}^2-5 r_{q^\prime}^2-3\right)+\left(r_{q^\prime}^2-1\right){}^2+4 r_V^4\Big)C_{q^\prime V}
\Big].
\end{align}

As far as the results for $q^2=0$ are concerned, they read
\begin{align}
\mathcal{V}_{qq'}^{V}(0)&=\frac{1}{r_V^2-\chi _{q^\prime}^2}\Big[8 r_V^6
-4 \left(r_{q^\prime} \left(3 r_{q^\prime}+2\right)+2\right) r_V^4+2 \left(r_{q^\prime} \left(r_{q^\prime} \left(2 r_{q^\prime}+7\right)+4\right)-5\right) r_V^2\nonumber\\&+2 \left(r_{q^\prime}^2-1\right){}^2 \left(2 r_{q^\prime} \chi _{q^\prime}+1\right)-\Big(4 \delta _{q^\prime}^2 r_{q^\prime} \chi _{q^\prime}^3+4 r_{q^\prime} \chi _{q^\prime}^2 r_V^2-4 \left(r_{q^\prime} \left(3 r_{q^\prime}+2\right)+3\right) r_V^4+8 r_V^6\Big)B_{q^\prime V}\nonumber\\& - \Big(4 \delta _{q^\prime} r_{q^\prime} \chi _{q^\prime}^2 r_V^2+4 \left(r_{q^\prime} \left(r_{q^\prime}+2\right)+3\right) r_V^4-8 r_V^6\Big)B_V + 4 r_{q^\prime} \left(\delta _{q^\prime}^2 \chi _{q^\prime}^3+r_{q^\prime} \chi _{q^\prime}^2 r_V^2-2 r_{q^\prime} r_V^4\right)B_{q^\prime}
\Big],
\end{align}
and
\begin{align}
\widetilde{\mathcal{D}}_{qq'}^{V}(0)&=
\frac{r_{q^\prime}}{(1-(r_{q^\prime}-r_V)^2)(1-(r_{q^\prime}+r_V)^2)}
\Big[
\left(r_{q^\prime}^2-r_V^2-1\right) \left(4 r_V^4-\left(5 r_{q^\prime}^2+3\right) r_V^2+\left(r_{q^\prime}^2-1\right){}^2\right)\nonumber\\& + \Big(4 r_{q^\prime}^2 r_V^4+\left(-5 r_{q^\prime}^4+4 r_{q^\prime}^2+1\right) r_V^2+\left(r_{q^\prime}^2-1\right){}^3\Big)B_{q^\prime} + \Big(\left(5 r_{q^\prime}^2+3\right) r_V^4-\left(r_{q^\prime}^2-1\right){}^2 r_V^2-4 r_V^6\Big)B_V\nonumber\\&-\Big(3 \left(3 r_{q^\prime}^2+1\right) r_V^4-6 r_{q^\prime}^2 \left(r_{q^\prime}^2-1\right) r_V^2+\left(r_{q^\prime}^2-1\right){}^3-4 r_V^6\Big)B_{q^\prime V}
\Big].
\end{align}

The loop functions of Eqs. \eqref{Sq2CMDM} and \eqref{Sq2CEDM} are given by
\begin{align}
\mathcal{P}_{qq'}^{S}(q^2)&=\frac{1}{\left(\hat{q}^2-4\right)^2}\Big[
\left(4-\hat{q}^2\right) \left(r_{q^\prime}^2-r_S^2+1\right)
+ \Big(\hat{q}^2 \left(2 r_{q^\prime}-1\right)+6 r_{q^\prime}^2-8 r_{q^\prime}-6 r_S^2-2\Big)B_{\hat{q}{q^\prime}}
\nonumber\\&+\Big(2 \delta _{q^\prime}^2 \chi _{q^\prime} \left(1-3 r_{q^\prime}-\hat{q}^2\right)+2 r_S^2 \left(\hat{q}^2 \left(r_{q^\prime}-2\right)+6 r_{q^\prime}^2-4 r_{q^\prime}+2\right)-6 r_S^4\Big)C_{q^\prime S}\nonumber\\& + \Big(\delta _{q^\prime} \left(\left(\hat{q}^2-10\right) r_{q^\prime}-\hat{q}^2-2\right)-\left(\hat{q}^2-10\right) r_S^2\Big)B_{q^\prime S} + \left(\hat{q}^2-4\right)\left(r_S^2B_S-r_{q^\prime}^2 B_{q^\prime}\right)
\Big],
\end{align}
and
\begin{align}
\mathcal{D}_{qq'}^{S}(q^2)&=\frac{ r_{q^\prime}}{\hat{q}^2-4}\Big[
B_{q^\prime S}-B_{\hat{q}{q^\prime}} +  \left(r_{q^\prime}^2-r_S^2-1\right)C_{q^\prime S}
\Big].
\end{align}

For $q^2=0$ we obtain
\begin{align}
\mathcal{P}_{qq'}^{S}(0)&=\frac{1}{2 \left(\chi _{q^\prime}-r_S\right) \left(r_{q^\prime}+\chi _S\right)}\Big[
\left(4 r_{q^\prime}^2+2 r_{q^\prime}-1\right) r_S^2-2 r_{q^\prime}^4-2 r_{q^\prime}^3+r_{q^\prime}^2-2 r_S^4-1\nonumber\\&+2 \left(\delta _{q^\prime} r_{q^\prime} \chi _{q^\prime}^2-r_{q^\prime} \left(2 r_{q^\prime}+1\right) r_S^2+r_S^4\right)B_{q^\prime S} + 2 r_{q^\prime} \left(r_{q^\prime} r_S^2-\delta _{q^\prime} \chi _{q^\prime}^2\right)B_{q^\prime}+ 2 r_S^2 \left(r_{q^\prime}^2+r_{q^\prime}-r_S^2\right)B_S
\Big],
\end{align}
and
\begin{align}
\mathcal{D}_{qq'}^{S}(0)&=\frac{1}{(1-(r_{q^\prime}-r_S)^2)(1-(r_{q^\prime}+r_S)^2)}\Big[
\left(1-r_{q^\prime}^2+r_S^2\right){}^2 +  \left(r_S^2 \left(1-r_{q^\prime}^2+r_S^2\right)\right)\left(B_S-B_{q^\prime S}\right)\nonumber\\&+
\left(2r_S^2-\left(1-r_{q^\prime}^2+r_S^2\right)\left(1-r_{q^\prime}^2\right)\right)\left(B_{q^\prime S}-B_{q^\prime}\right)
\Big].
\end{align}

\subsection{Two-point scalar functions}
In closing we present the closed form solutions for the two-point Passarino-Veltman scalar functions appearing in the calculation. The three-point scalar functions are too lengthy to be shown here.
\begin{align}
B_0(0,m_a^2,m_a^2)&=-\log\left(\frac{m_a^2}{\mu^2}\right)+\frac{1}{\epsilon}+\log(4\pi)-\gamma_E,\\
B_0(\hat{q}^2,m^2_{q^\prime},m^2_{q^\prime})&=
\frac{\sqrt{\hat{q}^2-4 r_{q^\prime}^2}}{\left|
 \hat{q}\right| }\log
   \left(\frac{\left| \hat{q}\right|
    \sqrt{\hat{q}^2-4 r_{q^\prime}^2}-\hat{q}^2+2
   r_{q^\prime}^2}{2
   r_{q^\prime}^2}\right)+2-\log
   \left(\frac{m_{q^\prime}^2}{\mu^2}\right)+\frac{1}{\epsilon}+\log(4\pi)-\gamma_E,\\
B_0(m_q^2,m_{q^\prime}^2,m_b^2)&=\sqrt{\lambda
   \left(x_q^2,x_b^2,x_{q^\prime}^2\right)}
   \log
   \left(\frac{\sqrt{\lambda
   \left(x_q^2,x_b^2,x_{q^\prime}^2\right)}+
   \left(r_b^2+r_{q^\prime}^2-1\right)}
   {2 r_b r_{q^\prime}}\right)
   +\frac{1 }{2}
   \left(1-r_b^2+r_{q^\prime}^2\right)
   \log
   \left(\frac{r_b^2}{r_{q^\prime}^2}\right)\nonumber\\&+2-\log
   \left(\frac{m_b^2}{\mu^2}\right)+\frac{1}{\epsilon}+\log(4\pi)-\gamma_E,
\end{align}
where $\lambda(x,y,z)=x^2+y^2+z^2-2(xy-xz-yz)$. The scale $\mu$ and the pole $\epsilon$ of dimensional regularization cancel out in the final result.

\bibliography{nuearticle}

\end{document}